%% file: samplefile.tex
\documentclass[11pt,twocolumn]{article}
\usepackage{pkg/preamble}

\usepackage{pkg/symbols}
\usepackage{authblk}
\title{This is some thing}
\author[ ]{Ahmed Taha}
\author[ ]{Spyros Boukoros}
\author[ ]{Jesus Luna}
\author[ ]{Stefan Katzenbeisser}
\author[ ]{Neeraj Suri}
\affil[ ]{Technical University of Darmstadt, Germany}
\affil[ ]{ \textit{\{ataha,jluna,suri\}@deeds.informatik.tu-darmstadt.de, \{boukoros,katzenbeisser\}@seceng.informatik.tu-darmstadt.de}}
\title{QRES: Quantitative Reasoning on  Encrypted Security SLAs}
\date{}

\begin{document}

\maketitle
\input{src/2-abstract1-N-A}
\input{Intro-A}

\input{src/5-background}

\input{src/4-Requirement}
\input{src/7-regtoken}
\input{src/1-tokentr}
\input{src/8-architecture}

\input{src/9-ranking}

\input{src/11-FormalAnalysis}
\input{src/10-Implementation}
\input{src/11-Relatedwork}

\input{src/12-conclusion}
\bibliographystyle{plain}
\bibliography{sigproc} 
\input{src/notation-table}
\input{src/13-app}

\end{document}

%% file: src/2-abstract1-N-A.tex

\begin{abstract}{
While regulators advocate for higher cloud transparency, many Cloud Service Providers (CSPs) often do not provide detailed
information regarding their security implementations in their Service Level Agreements (SLAs). 
In practice, 
CSPs are hesitant to release detailed information regarding their security posture for security and proprietary reasons.
This lack of transparency hinders the adoption of cloud computing by enterprises and individuals. Unless CSPs share information regarding the technical details of their security proceedings and standards, customers cannot verify which cloud provider matched their needs in terms of security and privacy guarantees. 
%
To address this problem, we propose QRES, the first system that enables (a) CSPs to disclose detailed information about their offered security services in an encrypted form to ensure data confidentiality, and (b) customers to assess the CSPs' offered security services and find those satisfying their security requirements. 
Our system preserves each party's privacy by leveraging a novel evaluation method based on Secure Two Party Computation (2PC) and Searchable Encryption techniques. 
We implement  QRES and  highlight its usefulness by applying it to existing standardized SLAs. 
The real world tests illustrate that the system runs in acceptable time for practical application even when used with a multitude of CSPs.
We formally prove the  security requirements of the proposed system against a strong realistic adversarial model, using an automated cryptographic protocol verifier.
}
\end{abstract}





%% file: Intro-A.tex

\section{Introduction}

Cloud computing allows customers to develop, manage, and access a spectrum of  resources (storage, software, applications, etc.) which are typically offered as-a-service in a remotely accessible fashion. 
In such a service-based environment, the cloud provisioning relies on stipulated Service Level Agreements (SLAs). Such an agreement is basically a contract between the Cloud Service Provider (CSP) and the customer regarding the offered service. 
These SLAs specify the cloud service levels requested by the customers, and required to be achieved by the CSPs.
A variety of parameters for different aspects of a service can be included in the SLA, such as but not limited to: availability, performance, downtime and location of the data. 
%

Albeit the numerous claimed benefits of the cloud to ensure confidentiality, integrity, and availability of the stored data, the number of security breaches is still on the rise \cite{secBr,Popa2011}. The lack of security assurance and transparency has prevented customers and enterprises from trusting the CSPs, and hence not using their services. Unless the customers security requirements are identified, documented, and communicated by the CSPs, customers can not be assured that the CSPs will satisfy their requirements. 

%

In this context, a number of cloud community stakeholders (e.g., ISO 27002 \cite{ISO} and the European Union Agency for Network and Information Security (ENISA) \cite{enisa11}) are pushing towards the inclusion of security parameters and CSP's security implementation in security SLAs (named secSLAs \cite{jesus2015}). Basically, secSLA has the same SLA structure however, it discloses detailed security-related information\footnote{Examples of security related information are ciphers used to encrypt data, vulnerability management/assessment procedures, minimum/average incident response times, security controls and configuration elements such as metrics for measuring cybersecurity performance, etc.} about each CSP security offers. The customers can use this information to assess and compare different service offerings provided by various CSPs and then select the best CSP that satisfies their requirements.


Despite the benefits of these kind of information, still CSPs do not disclose security related information in their secSLAs for security and/or commercial reasons \cite{ENISA}. The dangers of including security related information in the secSLA where pointed out by ENISA \cite{ENISA} as: 
\begin{enumerate}[-]
 \item Publicly disclosing security parameters may assist attackers to penetrate the system using a hole in the publicized data. Accordingly, the rate of malicious security breaches increases, which can be much harder to detect. This can also lead to a significant financial loss as a result of the customers compensation.
 \item Publicly detailing commercial sensitive information (i.e., financial terms, service levels, cost information, vulnerability descriptions which may include proprietary information, etc.) can be used by other competitors to improve their services.
\end{enumerate}

To that end, we tackle the aforementioned problem by designing and implementing a system called \textit{QRES} (Quantitative Reasoning on Encrypted SLAs). Our system simultaneously allows:
\begin{enumerate}
\item CSPs to specify their services along with the key measurable parameters in secSLAs, without revealing information about the offered security parameters or commercial sensitive information.
\item Customers to assess and evaluate the CSP's offered security services and choose the best CSP matching their needs.
\end{enumerate}

\begin{figure}[h]
\centering 
\scalebox{0.28}{\includegraphics{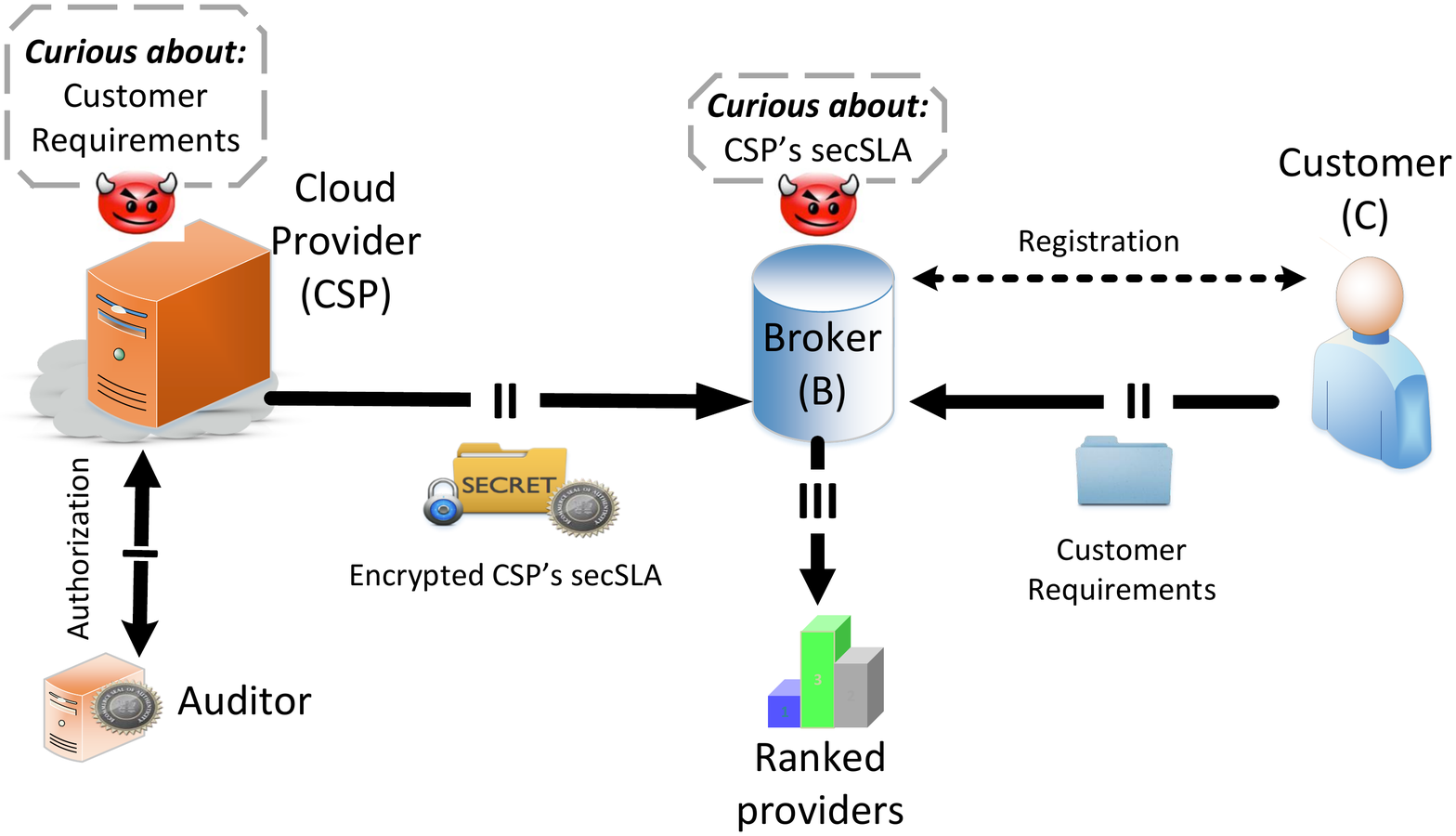}}
\caption{\textit{QRES} System Model. The system involves a customer (C) who wants to find the best CSP according to her/his needs. CSPs are certified by a trusted certification authority. A broker (B) receives the CSPs' encrypted secSLAs and the customer's requirements. Then, it ranks the CSPs according to the customer requirements and returns the result to the customer}
  \label{fig:RA}
\end{figure}

In our system (Figure \ref{fig:RA}), CSPs' encrypted secSLAs are certified and digitally signed by an trusted certification authority (i.e., auditor). The trust assumption relies on the fact that the CSPs' certificates are valid and trusted and thus, their encrypted secSLAs are verified and digitally signed by the auditor. After a successful authorization, every provider possesses a digital signature on their encrypted secSLA. The CSPs send their signed, encrypted secSLAs to an intermediate broker afterwards.  After the broker verifies the auditor's signature, it stores each encrypted secSLA  in a database.

Furthermore, customers send their requirements to the broker\footnote{In this paper, we assume the case of a novice or basic customer who can not search for her/his over encrypted data} which tries to match  the customer's requirements against the stored encrypted data in order to find the best matching CSP's secSLA. During this phase, neither the broker can learn the $\p$'s encryption key nor the CSP's can learn the customer's requirements. Finally, the broker sends the customer the CSPs ranking according to the customer requirements. 
\newline

To verify the \textit{QRES} system's correctness\footnote{\textit{QRES} must ensure that the entities should learn nothing except their output and each entity should receive its correct output.}, we start by formally defining the required security properties; the presented system must ensure the following privacy requirements: data confidentiality and data fairness.
Then, we conduct a formal security analysis of \textit{QRES} using ProVerif~\cite{blanchet2001efficient}, an automated cryptographic protocol verifier, establishing the defined properties against a strong adversarial model (i.e., Dolev-Yao adversary model \cite{dolev1983security}).
To validate our system model, we implement \textit{QRES} using Amazon AWS DynamoDB \cite{ddb}. We utilize real-world CSPs' secSLAs found on the public STAR (Security, Trust and Assurance Registry) \cite{csa_star} repository, which are complaint with the relevant ISO/IEC 19086 standard \cite{iso19086}.

The ProVerif scripts and proofs used in this paper as well as the system implementation are publicly available at \cite{qres}.

\noindent
\textbf{Contributions.}\hspace{0.25cm} The contributions of our system are summarized below.
\begin{enumerate}
\item We propose the first  system (\textit{QRES}) which enables  CSPs to publicly  disclose detailed information about their offered services in encrypted secSLAs and customers to assess the CSPs' offered security services and find those satisfying their requirements. 
\item \textit{QRES} is built around a novel two-party privacy preserving query over encrypted data scheme (named \textit{QeSe}).
\item We  formally prove the  system's correctness by defining the needed security properties and proving that the system holds these properties against a strong adversarial model.
\item We implement,  evaluate and benchmark \textit{QRES} using real-world CSPs' secSLAs. We show that the system's performance is practical for the presented use case.
\end{enumerate}

\subsection*{{Outline}}
The rest of the paper is organized as follows: The basic concepts of the cryptographic tools, notations, and security definitions used are developed in Section \ref{sec:notation}. Section \ref{sec:RA} defines the system and threat models, as well as the system objectives and requirements.  Section \ref{sec:SA} elaborates the architecture of the proposed system. 
A security analysis of the \textit{QRES} system is presented in Section \ref{sec:SC}.
The implementation of the presented system is detailed in Section \ref{sec:imp}.  
We present the related work in Section 7 and the conclusion in Section \ref{sec:conc}.

%% file: src/5-background.tex

\section{Basic Concepts}
\label{sec:notation}
This section briefly explains the basic concepts, cryptographic tools, notations and security definitions used in this paper. We adapt and implement well known cryptographic mechanisms, and therefore we do not include proofs as they already exist in the referenced papers. 

\subsection{Notations and Preliminaries}
\label{sec:def}
We summarize most of the terminology used in Table \ref{table:terms} (in the Appendix). Furthermore, we define the searchable encryption notations used in the paper in the Appendix.

\subsection{Security Service Level Agreements}
\label{sec:SLA}
A security  service level agreement  describes the CSP's offered security services, and represents the binding commitment between a CSP and a customer. Basically, each CSP's  secSLA consists of a number of offered security services which contain a list of Service Level Objectives (SLOs). The SLOs are the single measurable elements of an  SLA/secSLA that specify the cloud service levels required by the customers and to be achieved by the CSP. Each SLO is assessed using one or more key measurable parameters. These parameters help in the measurement of the cloud service objectives by defining measurement rules that facilitate the assessment and decision making. 

For example, how a CSP recovers from incidents is typically defined in terms of severity and time to recovery. As specified by ENISA~\cite{ENISA}, a severity classification scheme detailing levels from 1 to 5 could be defined in an  SLA, where a ``N/A'' level could be included in the  SLA for incidents which have no security impact. The criteria of each level is based on various parameters. 

\begin{figure}[h]
\centering 
\scalebox{0.3}{\includegraphics{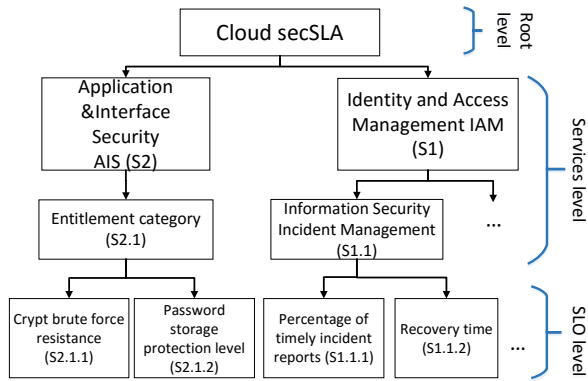}}
\squeezeup
\squeezeup
\caption{Cloud  secSLA hierarchy based on security posture provided by the STAR repository and compliant with the relevant ISO/IEC 19086 standard.}
  \label{fig:SLA1}
\end{figure}

Based on the analysis of the state of practice presented in \cite{jesus2015}, security  SLAs are modeled using a hierarchical structure, as shown in Figure \ref{fig:SLA1}. The root of the structure defines the main container for the  secSLA. The intermediate levels are the services which form 
the main link to the CSP's offered services. The lowest level (SLO level) 
represents the actual SLOs committed by the CSP and consequently offered to the cloud customer. To formalize the concept of an SLA/secSLA, we use the definition presented in \cite{taha16}.

\begin{definition}
An  SLA consists of a set of services $S$. Each service consists of a finite positive number $n$ of SLOs $o_i$; where $i = 1 \ldots n$. Each SLO consists of $l$ different values $v$; such that $o_i = {v_{1}, v_{2} , \ldots , v_{l}}$. Each of these values implies a specific service level offered by the CSP and required by the customer.
\end{definition}

We illustrate the secSLA's structure and functionality better with an example.
We consider a customer processing financial transactions using Software-as-a-Service (SaaS). The customer looks for a secSLA which specifies a recovery time objective of less than 1 minute and a monthly report offered by the selected CSP specifying the mean recovery times~\cite{ENISA}. Using this example we specify two SLOs (``Percentage of timely incident reports $S_{1.1.1}$'' and ``Recovery time $S_{1.1.2}$'') as shown in Figure \ref{fig:SLA1}. The ``Percentage of timely incident reports'' SLO is composed of  \{$yearly, half-yearly, monthly, weekly$\} values which are defined using service levels as $level_1, level_2, \ldots, level_4$ respectively. If a CSP is committing a ``Percentage of timely incident reports'' of $monthly$, then $v_{S_{1.1.1}}= level_3$. Similarly, a CSP commits other SLOs so that the overall CSP's  secSLA contains a list of SLOs with different values that is committed to fulfil. 
 
%
%

\begin{lstlisting}[language=XML,caption=Excerpt of the  SLA depicted in Figure \ref{fig:SLA1},captionpos=b]
< SLA slaid=``sla1">
<service id=``S1" name =``Identity and Access Management" category=``IAM" pre=``1">
<control id=``S1.1" name =``Information Security Incident Management" category=``IAM-09" pre=``2">
<slo id=``S1.1.1" name =``Percentage of timely incident reports" value=``level3'' pre=``3"></slo>
<slo id=``S1.1.2" name =``Recovery time" value=``level2" pre=``4"'></slo>	
</control></service></ SLA>
\end{lstlisting}

\noindent
\textbf{Format of secSLAs.}\hspace{0.25cm} The  secSLA  shown in Figure \ref{fig:SLA1} can be specified using a machine readable format such as an XML structure as depicted in Listing 1.
The XML data values can be captured using the pre-fields shown in Listing 1 (named ``pre''), which are sequence numbers that count the open tags in an XML. An example of how the pre-fields are computed is depicted in Table \ref{table:xml}. Each pre-field is used as a service identifier as each service/SLO has a unique pre-field in the  secSLA.

\begin{table}[h]
\centering

\begin{tabular}{|l|c|}
\hline
\textbf{XML structure}&\textbf{pre-field}\\\hline
\hline
$<S_1>$&1\\\hline
\hspace{0.3cm}     $<S_{1.1}>$&2\\\hline
\hspace{0.6cm}            $<S_{1.1.1}$ value=$level_3>$ &3\\\hline
\hspace{0.6cm}           $<S_{1.1.2}$ value=$level_2>$ &4\\\hline
\hspace{0.3cm}      $</S_{1.1}>$&\\\hline

\end{tabular}
\caption{Excerpt of an  SLA XML structure with the calculation of pre-fields}
\label{table:xml}
\end{table}

Note that, customers can only assess the CSPs and choose the best one satisfying their  requirements, only if the CSPs' shared  information are relevant to customers' concerns (i.e., stemmed from customers' requirements) \cite{cyber}. To achieve this, the customer   $\cust$ has to create her/his set of  requirements using the same secSLA-XML structure used by the CSP.

\subsection{Searching Over Encrypted Data}
\label{sec:SSE}
We define the problem of searching over encrypted data using the following example. Assume a customer who encrypts her/his documents and stores those at the CSP's storage server. However, by encrypting these documents, the customer can not search for certain keywords anymore and thus,  content retrieving is very inefficient. The basic approach for retrieving the required data related to a certain keyword would require the customer to download \textit{all}  stored encrypted documents, and then decrypt them to perform the keyword search. However, this solution is  time consuming and impractical. In addition, retrieving all files   incurs unnecessary network traffic, which is undesirable in the pay-as-you-use cloud paradigm used today.

From the above problems,  the need arises for an efficient data retrieval  scheme which enables the customer to search directly over encrypted data.  A  solution to this problems is what is widely known as searchable encryption (SE). 
SE allows a customer to encrypt data in such a way that she/he can later generate search tokens to send queries to the CSP. Given these tokens, the CSP can search over the encrypted data and retrieve the required encrypted files. As specified in \cite{kamara2012dynamic}, a SE scheme is secure
if: (i) the ciphertext alone reveal no information about the encrypted data, (ii) the encrypted data together with a search token (i.e., queries) reveals at most the result of the search, and (iii) search tokens can only be generated using the same encryption key used to encrypt the data. 

There exists a large number of SE schemes \cite{song2000practical, boneh2004public, bellare2007deterministic, kamara2012dynamic} which are either deterministic or randomized. Deterministic schemes \cite{bellare2007deterministic} encrypt the same message to the same ciphertext. However, it does not protect against frequency analysis attacks. On the other hand, randomized schemes \cite{song2000practical, kamara2012dynamic} prevent frequency analysis by salting ciphertexts and thus providing stronger security guarantees. However, the usage of  salt in these schemes requires combining each token with each salt, resulting in a processing time linear in the number of salts for each token. In \cite{sherry2015blindbox} Sherry et al., introduced an encryption scheme which achieves both the detection speed of deterministic encryption and the security of randomized encryption. Our system uses a deterministic encryption scheme for searchable encryption.

\subsection{Privacy Preserving Computations}
\label{sec:SC}
\textbf{Secure two party computation.}\hspace{0.25cm}
The aim of secure two-party computation is to enable both parties to carry out computing tasks without revealing information of any kind about private data to the participants. Assume two parties, $A$ and $B$ have some private information. They want to learn the result of some function using both of their inputs, while each party would learn nothing about the other party input. To achieve this, Yao's protocol based on garbled circuit (named Yao garbled circuit) \cite{yao1986generate,lindell2009proof} is used.  Yao's protocol based on garbled circuits allow two parties to exchange a garbled circuit and garbled inputs for a function, which can be used to 
compute an output without leaking information about their inputs.

A garbled circuit is a circuit that consists of garbled gates and their decryption tables. In a garbled gate, two random bits have been selected  for every input wire to the gate, representing 0 and 1. Those bits garble the gate, making it impossible to compute the output unless someone has access to the garbled computation table. The garbled computation table maps essentially the random inputs to the output of the gate, which is also random.

We illustrate the basic idea of how Yao's garbled circuit can be used, assume  two parties $A$ and $B$ that have  secret inputs, $x$ and $y$ respectively. Both  of them want to compute $F(x,y)$ without revealing their private inputs to each other ($x$ to $B$ and $y$ to $A$). 
To achieve this, one party (for instance $A$) prepares a garbled version of the computing function $F$ (named $\garble F$). Basically, $\garble F$ produces the same output as $F(x,y)$, if given the corresponding encoding of each bit of both inputs $x$ and $y$.

The inputs garbling is done by producing a pair of labels for each input bit of $F$ (${G}^0,{G}^1$, that is one label corresponds to bit $0$ and the other to $1$). 
Next, $A$ sends $\garble F$ along with the encoding of $x$ to $B$, which only needs the encoding of $y$ from $A$ to compute the $\garble F$ without learning any intermediate values. 
For this task, both parties use oblivious transfer \cite{rabin2005exchange,naor1999oblivious,asharov2013more}.


\textbf{Oblivious transfer (OT).}\hspace{0.25cm} OT is a crucial component of the garbled circuit approach, as it enables party $B$  to obtain the encoding of the $b$ bit from $A$, without (i) $A$ knowing $b$ and (ii) $B$ learning the encoding scheme. In this way, party B can request from A the keys that he can use for his input encoding without i) A learning   B' input and ii) B exploiting the protocol by having access to the encoding scheme and computing much more than allowed.

%% file: src/4-Requirement.tex
\section{Requirements Analysis}
\label{sec:RA}
In this section, we describe the system model, present the system requirements, and define our threat and trust models.

\subsection{System Overview}
\label{sec:SM}
Finding the best matching CSP (according to the customer's security requirements) is the objective of the proposed system. Our system model (depicted in Figure \ref{fig:RA}) involves $m$ $\p$s\footnote{Throughout the paper, we explain our model and queries searching scheme using only one $\p$. Nevertheless, the same model applies for all $\p$s.}, a customer $\cust$, and a broker $\broker$. 
The customer $\cust$ is a company or an individual who is searching for the best provider that satisfies her/his requirements. The $\p$s are cloud providers that disclose information about the offered security posture in their secSLAs. $\p$s are encrypting their secSLAs before sending them to the broker. The broker $\broker$ is an entity that performs the searching of the customer requirements over the $\p$s' encrypted secSLAs on behalf of the customer. Hence, $\broker$ ranks and manages the selection of the best matching $\p$.

\subsection{Threat Model}
\label{sec:TM}
Security literature distinguishes between two adversarial model for secure computation; participants can be either semi-honest or malicious. In this work we consider a semi-honest (also known as honest-but-curious) threat model. This is a commonly used security model for secure computation (we refer the reader to Goldreich \cite{goldreich2009foundations} for details) where the semi-honest participants correctly follow the introduced protocol but  attempt to obtain additional information about the other participants. By considering the semi-honest model, a dishonest participant observing the system's network should not be able to alter or recover stored data.

The semi-honest setting is relevant in this study, as all entities, the $\p$, $\broker$, and $\cust$, would like to continue the protocol; acquire the best matching $\p$ according to \cust's requirements. 
However, each entity can attempt to obtain additional information about the other entity's input as depicted in Table \ref{table:threat}. 
\begin{table}[h]
\centering
  \begin{tabular}{|p{1.4cm}|p{5cm} |}
  \hline
 \textbf{Customer} \cust &  Tries to learn the $\p$'s private key $\key$ in order to learn the $\p$'s secSLA \\\hline
  \textbf{Cloud Provider} $\p$ & Tries to learn \cust's requirements
  or sends a faulty secSLA\footnote{Faulty secSLAs are secSLAs which contain service levels the $\p$ does not offer, and could not fulfil, in order to match the customer requirements} \\\hline
  \textbf{Broker} \broker & Tries to (a) alter \cust's requirements in order to match a colluded $\p$, or (b) collude with $\cust$ to learn the $\p$s' secSLAs and identify their identities \\\hline
  \end{tabular}
  \caption{The semi-honest threat model of our system. Every entity tries to enhance their knowledge about the others while following the protocol.}
     \label{table:threat}    
\end{table}

\subsection{\textbf{Trust Model}}
\label{sec:trust}
In this paper we consider a trusted auditor.
Before detailing the proposed system model, we note that customers can only trust the result of an assessment if the information taken as an input is reliable. In other words, in order to guarantee the validity of the proposed system, the encrypted secSLAs provided by the participating $\p$s are required to be certified from the trusted certification authority (the auditor). 
For example, an auditor certifying the CSP's security posture as reflected by its secSLA (e.g., based on an ISO 27001 certificate \cite{csa_ocf}).
The trust assumption relies on the fact that the $\p$s' certificates are valid and trusted and thus, their encrypted secSLAs are verified and digitally signed. Such signing scheme should provide a proving statement without revealing the secSLA input. For simplicity, we assume that the participating $\p$'s encrypted secSLAs are to be verified and digitally signed. We summarize the initial knowledge of each entity in Table \ref{table:trust}.

\begin{table}[h]
\centering
   
  \begin{tabular}{|p{2.6cm}| l | c | c | c | }
  \hline
  \multirow{2}{*}{Parameter} &\multicolumn{4}{ c |}{Each entity's knowledge}\\\cline{2-5}
  &{\cust} &{\p} & {\broker} & {Auditor} \\\hline
\hline  
  \p's secSLA & &$\times$ &&\\\hline
  \p's Encryption Key & &$\times$ &&\\\hline
  \p's Certificate/ID & &$\times$ &&$\times$\\\hline
  \cust's Requirements &$\times$& &$\times$&\\\hline
  \end{tabular}
    \caption{In the table we summarize every entity's initial knowledge. Every participating entity has  minimal knowledge regarding the others.}
     \label{table:trust}    
\end{table}

\subsection{System Requirements}
\label{sec:Req}
In order to provide the privacy and correctness guarantees, the presented system must ensure: 
\begin{enumerate}[1)]

\item \textit{Input Validation:} The $\p$s' encrypted secSLAs provided by the participating $\p$s are  digitally signed by an auditor.

\item \textit{Fairness:} The system must ensure that if one entity ($\p$ or $\broker$) quits the computation, it can not learn more information  than the other entity. In other words, none of the entities can learn the result first and then abort.

\item \textit{Data Confidentiality:} 
The encrypted $\p$'s secSLA can only be decrypted using its $\p$'s private key $\key$.  Further,  the broker $\broker$ should  \textit{only} learn the output of the searching queries (i.e., the search queries represent the customer requirements). Moreover, the $\p$ should not have access to $\cust$'s requirements to avoid changing its secSLA specifications according to those.

Note that, $\broker$ or $\cust$ can ensure the secSLA compliance by monitoring and verifying the offered service levels by (i) using appropriate log samples provided by the $\p$, and/or (ii) adding alerts and triggers based on the service key measurable parameters \cite{ENISA}. 
However, in this paper we only focus on the CSPs evaluation and services assessment and refer the readers to \cite{ENISA, US} for further monitoring process details. 

%% file: src/7-regtoken.tex
\begin{figure*}[h]
\centering 
\scalebox{0.25}{\includegraphics{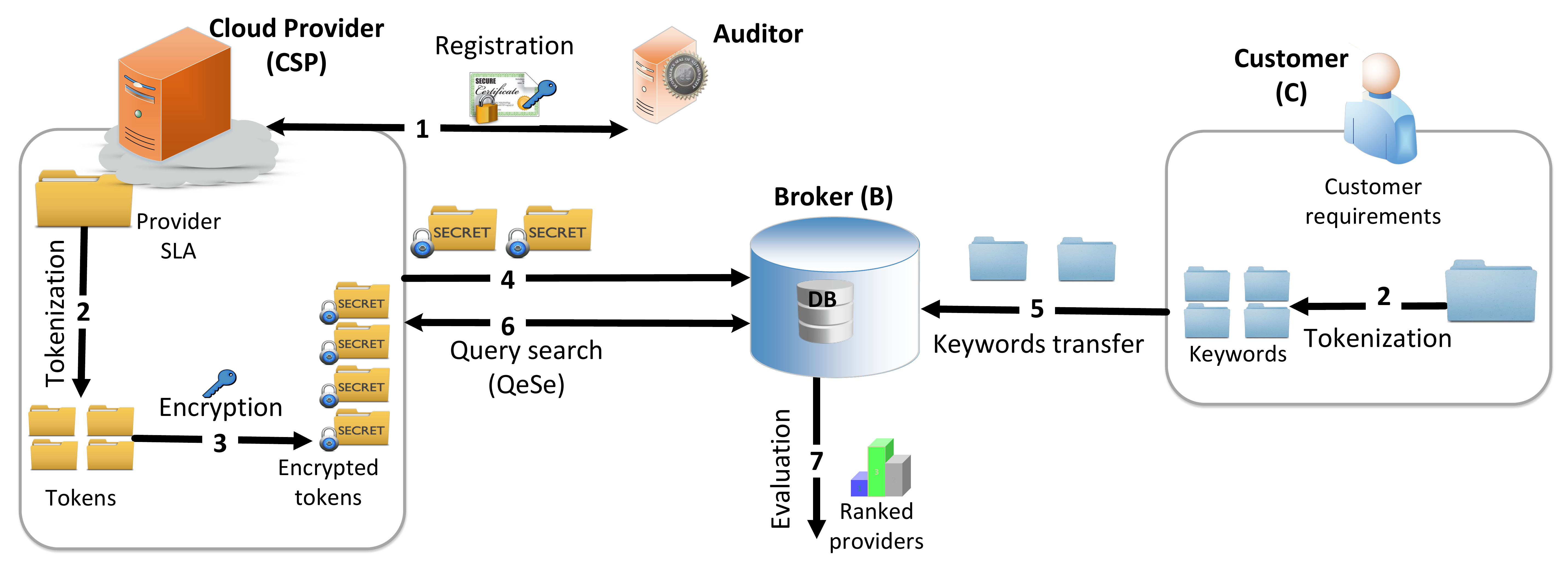}}
\caption{\textit{QRES} System Architecture. Every step is  ordered with an increasing number. The procedure in our system begins with an auditor verifying the cloud providers. Then continues with the tokenization of the secSLAs and the customers' requirements and ends with the \textit{QESE} protocol. This protocol enable the matching between the requirements and the secSLAs in the encrypted domain.}
  \label{fig:SM}
\end{figure*}

\section{QRES Architecture}
\label{sec:SA}

In this section, we detail each of  \textit{QRES} phases using the  progressive stages  depicted in Figure \ref{fig:SM}. As stated earlier, as an initial phase the customers register and are verified by the broker $\broker$ before \textit{paying} the broker for the offered assessment service. Furthermore, we assume that the broker's certificate ($\certb$) is validated by the auditor. Then, send by the auditor to the CSPs.

\subsection*{Stage 1: Providers Authorization.} Every CSP is first  registered and verified by a trusted auditor. The auditor verifies the CSPs certificate and then signs the encrypted secSLA. The broker B  can verify the validity of the encrypted secSLAs by checking for the auditor's signature.

\subsection*{Stage 2: Tokenization.}
\label{sec:Token}
After $\cust$ creates her/his set of  requirements using the same secSLA-XML structure as used by the $\p$  (similar to the template shown in Listing 1)\footnote{Cloud Security Alliance (CSA) has created a consensus assessments initiative questionnaire (CAIQ) \cite{csa_caiq}, to define the  controls contained in an SLA. $\p$s have used it as an SLA template by detailing their offered  controls and publishing them on the STAR repository \cite{csa_star}.}, both the $\p$ and $\cust$ tokenize their secSLA-XML data specifications such that: 
\begin{itemize}

\item Each $\p$ splits its   specified services (in the secSLA) into substrings. For every substring, it creates a fixed length token ( 8 bytes per token as depicted in Table \ref{table:terms}). The  generated tokens are denoted as ``$t_{\p}$'', such that $T_{\p}={t^1_{\p},\ldots, t^n_{\p}}$; where $T_{\p}$ specifies the set of services offered by the $\p$ in its secSLA and $n$ is the number of tokens.
\item Similarly, $\cust$ splits her/his requirements into substrings and then generates a fixed 8 bytes token for every substring. The customer's generated tokens are named ``keywords" and denoted as ``$w_{\cust}$'' so that $W_{\cust}={w_{\cust}^1, \ldots,w_{\cust}^n}$; where  $W_{\cust}$ is the set of customer requirements.  
\end{itemize}

Both the customer  and the $\p$ use the same secSLA template with the same services/SLOs where each of them specify different SLO values. Tokens are generated for each SLO, by searching for the SLO id and extracting only its value and pre-field of the specified SLO (i.e., each SLO in an SLA XML has a unique pre-field as specified earlier and depicted in Table \ref{table:xml}). For example, the tokens generated from Listing 1 are: ``$level3 || 3$'' and ``$level2 || 4$''.  
Therefore, for each secSLA no similar tokens can be generated as no equal pre-fields exist. Even if the $\p$ is offering two values for a specific SLO, both tokens would be different because of the different values.

%
%
%

%% file: src/1-tokentr.tex


\begin{figure*}
\begin{center}
\fbox{
\pseudocode{
\textbf{Provider} (\p) \< \< \textbf{Broker} (\broker)  \\[0.1\baselineskip]
 {knowledge: } \<\<{knowledge: }\\
\sk_{\p},\pk_{\p},\pk_{\broker}, \kpvauth,t^1_\p,\ldots, t^n_\p \< \<\sk_{\broker},\pk_{\broker}, \kpvauth, w^1_\cust,\ldots, w^n_\cust \\[0.1\baselineskip][\hline]
 \<\< \\[-0.5\baselineskip]
 \textbf{[1]} \hspace{0.1cm} \key_{\garble} \sample \kgen (\secparam)\<\<\\
 \textbf{[2]} \hspace{0.1cm} \gar_\key\leftarrow \enc(\key,\key_{\garble})\<\<\\ 
\textbf{[3]}\hspace{0.1cm} \text{Using \textit{OT:} } \<\< \textbf{[3]} \hspace{0.1cm} \text{Using \textit{OT:} } \\
G^{w^{1}_{{\cust}}} \leftarrow w^{1}_{{\cust}}  \< \sendmessageleft*[2cm]{\sig(\sk_{\broker},{w^{1}_{{\cust}}})}\<w^{1}_{{\cust}} \coloneqq w^{1}_{0} \ldots w^{1}_{n}\\
\text{where} \hspace{0.1cm} G^{w^{1}_{{\cust}}} \coloneqq G^{w^{1}_{{0}}} \ldots G^{w^{1}_{{n}}}\<\<\\
 \hspace{0.1cm} \sig(\sk_{\broker},G^{w^{1}_{{\cust}}}) \leftarrow \sig(\sk_{\broker},{w^{1}_{{\cust}}})\<\sendmessageright{top=$\gar_\key$ , bottom=${\sig(\sk_{\broker},G^{w^{1}_{{\cust}}})}$}\< \\
\< \< \textbf{[4]} \hspace{0.1cm} \vrfy(\pk_{\broker},\sig(\sk_{\broker},G^{w^{1}_{{\cust}}}) \text{ valid?}\\
 \< \< \textbf{[5]} \hspace{0.1cm} c^{G^{w^{1}_{\cust}}}\sample \gar_\key(G^{w^{1}_{{\cust}}})\\  
  \< \< \hspace{0.1cm} c^{G^{w^{1}_{\cust}}} \coloneqq \enc(\key, w^{1}_{{\cust}}) \\
\< \< \textbf{[6} \hspace{0.1cm} d \sample \search(c^{w^1_{\cust}},c^{t^1}_{{\p}} \ldots  c^{t^n}_{{\p}})\\
}
}
\end{center}
\caption{Secure computation protocol  between a CSP and the broker (QeSe). The initial knowledge and the generation of the garbled circuit are depicted in lines 1 and 2. The oblivious transfer start in line 3 and the secure two party computation protocol ends at line 5. Then, the broker has acquired the 
$\enc(\key, w^{1}_{{\cust}})$ and can match the requirements to the secSLAs. }
  \label{fig:token1}
\end{figure*}

%% file: src/8-architecture.tex

\subsection*{Stage 3: Tokens Encryption.}
\textit{QRES} utilizes a deterministic encryption scheme $\enc(\key,x)$, such that the encryption of the $\p$ token ($t_{\p}$) is denoted by $\enc(\key,t_{\p})$. For instance, in order to check if the $\p$'s token ($t_{\p}$) is matching the customer keyword ($w_{\cust}$), we can simply check if $\enc(\key,t_{\p})$ is equal to $\enc(\key,{w_{\cust}})$. Unfortunately, deterministic encryption schemes, which are rather fast,  cannot be used in every case, as every occurrence of $t_{\p}$ will result to the same ciphertext.
However, this is not the case in our system as every generated token $t_{\p}$ is unique by design  in the $\p$'s secSLA, as every token contains different pre-field. Therefore, we utilize the $AES-CBC$ encryption scheme to encrypt each $\p$'s generated token $AES_{\key}(t_{\p})$. We use a typical instantiation of a hash function, which is $\SHA256$, to compute the  initialization vector (refer to Table \ref{table:terms} for the hash function definition).

\subsection*{Stage 4: Tokens Transfer.}

Both the CSP and the customer, send their tokens to the broker.
$\broker$ verifies the auditor's signature and then saves the $\p$'s encrypted tokens in a database.
For $m$ $\p$s, $\broker$ saves $m$ different lists with encrypted tokens. 

\subsection*{Stage 5: Keywords Transfer.}
 
In this stage, $\broker$ receives $\cust$'s requirements (i.e., keywords) securely via an encrypted channel (e.g., SSL). 


%
%
Once $\broker$ receives $\cust$'s messages, 
it saves the messages and starts a secure two-party computation with each $\p$ in order to find the best matching $\p$ according to $\cust$'s security requirements in the next stage.

\begin{figure}[h]
\centering 
\scalebox{0.28}{\includegraphics{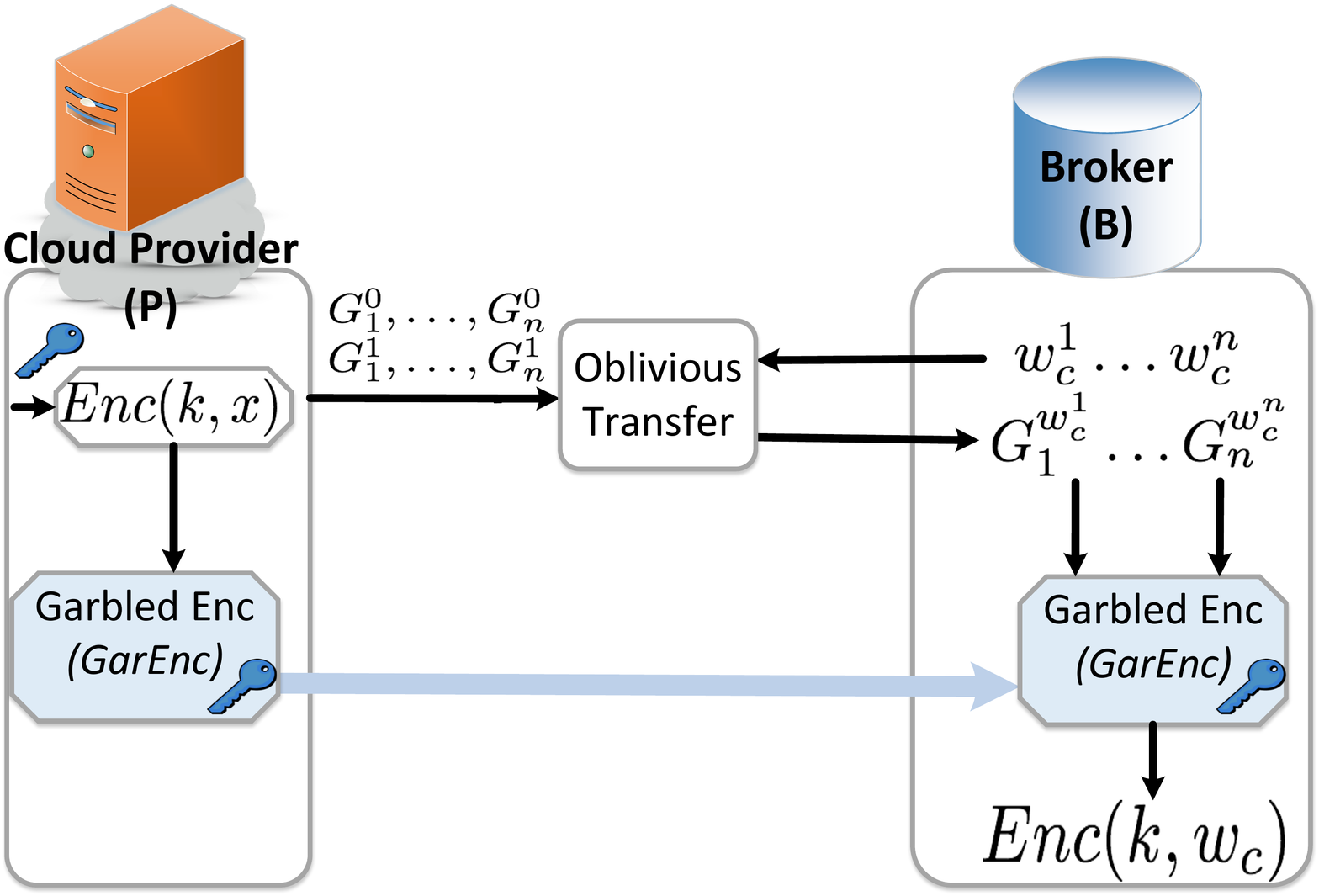}}
\caption{Illustration of the  Secure Computation Protocol between the CSP and the broker (\textit{QeSe}).}
  \label{fig:gar}
\end{figure}

\subsection*{Stage 6: Query Search (QeSe).}

\noindent
In order to design and build such a privacy preserving system, several challenges need to be addressed:

\begin{enumerate}[-]
\item In order for customers to search over the encrypted secSLAs, they must have access to the same secret keys as the CSPs. Consequently, the system should uphold the privacy of all parties by allowing  customers to search/match their requirements blindly without neither the CSPs learning the customers' queries, nor the broker (or customer) obtaining the CSP's secret key. 

 \item In addition, the system should  prevent malicious CSPs from deliberately providing false information trying to match the customer requirements.

\item Lastly, the system should prevent malicious customers from just gathering information about each CSP offered services and then quitting (i.e., by performing several searching iterations, each iteration with different requirements or by performing frequency analysis). 

\end{enumerate}

To tackle these problems, \textit{QRES} utilizes a novel 
two-party privacy preserving query over an encrypted data scheme (named \textit{QeSe}).
\textit{QeSe} is based on  two-party privacy preserving computation (i.e., Yao garbled circuit \cite{yao1986generate,lindell2009proof}) to allow  CSPs to encrypt the customers' search queries without learning them. 
Accordingly,  customers can find the best matching CSP's secSLA without knowing the CSP's encryption key.
Furthermore, the protocol supports a new security property by enhancing the security setting of the traditional two-party privacy preserving computation (Yao garbled circuit evaluation). 
This is achieved by validating the participants' inputs (i.e., CSP's secSLA and broker's signature on the customer's keywords) before computation.

%

\vspace{0.5cm}
\noindent
\textbf{\textit{QeSe}.}\hspace{0.25cm}
For simplicity, we  explain this stage using  one $\p$ and $\broker$. 
The broker  $\broker$  has already stored each $\p$ encrypted tokens ($\enc(\key,t_{\p})$) (from the previous steps) in a database. 
Furthermore, $\broker$ has already received the customer keywords.
In order to find the encrypted tokens matching the customer keywords, they have to be encrypted by the same encryption key used by $\p$  to encrypt the tokens. 
In other words, $\broker$ has to compute $\enc(\key,w_{\cust})$ for every $w_{\cust}$ using the same encryption key used by the $\p$. The main challenge here is for  $\broker$ to obtain the encrypted $w_{\cust}$ without knowing the $\p$'s secret key $\key$ and without allowing the $\p$ to learn $w_{\cust}$.

\textit{QeSe} allows all parties to jointly achieve their purpose (exchange k so $w_{\cust}$ can be encrypted) by running a  secure two party computation between the CSP and the broker. 
We enhance the security setting of the traditional two-party privacy preserving computation by allowing the garbled circuit to also check  the validity of both the encrypted tokens and the broker's digital signature of the  customer's requirements. In case there is a problem with the validation our protocol  quits.

The $\p$ provides $\broker$ with a ``garble'' of the encryption function with the key $\key$ hardcoded in it, as depicted in Figure \ref{fig:token1} (line 2) and Figure \ref{fig:gar} (we denote this garbled function by $\gar_{\key}$). This garbling hides the $\p$'s secret key $\key$. 
$\gar_{\key}(x)$ produces the same output of $\enc(\key,x)$ if given the corresponding encoding of each bit of input $x$ (i.e., ${G^x}$).
Thus, $\broker$ can run this garbled function on each keyword $w_{\cust}$ in order to  obtain the $\enc(\key,w_{\cust})$ only if he is able to acquire from the $\p$ an encoding for the $w_{\cust}$ ($G^{w_{\cust}}$) as depicted in line 3 in Figure \ref{fig:token1}. 
For this task, both parties ($\p$ and $\broker$) use an oblivious transfer protocol, where $\broker$ receives an encoding of $w_{\cust}$ from the $\p$ without revealing $w_{\cust}$. 
The encoding of keywords $w^{1}_{\cust}, \ldots, w^{n}_{\cust}$ is denoted as $G_1^{w^{1}_{{\cust}}}, \ldots, G_n^{w^{n}_{\cust}}$. 
Similarly, the broker  receives the encoding of $\sig(\sk_{\verifier},G_1^{w^{1}_{{\cust}}})$, that is the encoding of the keyword's signature (line 3 in Figure \ref{fig:token1}).
Note that, on input $\sig(\sk_{\broker},w_{\cust})$, $\gar_{\key}$ first checks if $\sig(\sk_{\broker},w_{\cust})$ is a valid signature of $w_{\cust}$ using $\broker$'s public key ($\pk_{\broker}$). 
If it is valid, $w_{\cust}$ is encrypted as $\enc(\key,w_{\cust})$.
It is important for the security of the protocol to mention  that,  a garbled circuit is no longer secure if $\broker$ receives more than one encoding for the same circuit. Thus, $\broker$ obtains a fresh, re-encrypted garbled circuit $\gar_\key$ for every $w_{\cust}$.
At the end of this process, $\broker$ gets the $\enc(\key,w_{\cust})$ for every $w_{\cust}$. Afterwards, $\broker$ searches for an exact matching of $\enc(\key,w_{\cust})$ over all $\p$s encrypted tokens (as depicted in line 6 in Figure \ref{fig:token1}). Each successful matching result is saved in a list at $\broker$ using an index $d$. The process is repeated for all $\p$s. Hence, for $m$ $\p$s, $\broker$ creates $m$ lists. Each of these lists contains the number of encrypted tokens matching $\cust$'s keywords. 
%

%% file: src/9-ranking.tex

\subsection*{Stage 7: CSPs Ranking.} 

As a last step, the $\p$s are evaluated and ranked according to the number of tokens matching the customer requirements. 
%
The $\p$ with the highest  number of encrypted tokens matching the customer keywords, is given the highest score and selected as the best matching provider. 

%% file: src/11-FormalAnalysis.tex
\section{Security Analysis}
\label{sec:SC}
In this section we formally analyse and verify   the security considerations of the system architecture described in Section \ref{sec:SA}, with respect to the security requirements defined in Section \ref{sec:Req}.

\subsection{Formal Analysis}
\label{sec:FA}

Formal methods can be used to model a cryptographic protocol and its security properties against an adversarial model, together with an efficient procedure to determine whether a model  satisfies those properties.

%
%

In this section, we analyse the security of the proposed system protocols with respect to the security objectives outlined in Section \ref{sec:Req}  using ProVerif \cite{blanchet2001efficient}. ProVerif is an automated verification tool that can handle an unbounded number of protocol sessions. It is used to model cryptographic protocols and their security properties against a strong adversarial model. In contrast to other state-of-the-art tools (e.g., the Avispa tools \cite{armando2005avispa} and Scyther \cite{cremers2008scyther}), ProVerif provides a larger feature set \cite{cremers2009comparing}. Furthermore, it allows the modelling of  cryptographic primitives using equational theory which is used to model the Yao's garbled circuit protocol.  

In order to verify  a system's  security, first we define a list of security properties. Then, the context in which the system functions are created. The system's context consists of assumptions about the environment and the adversarial model. We model the formal specification of the \textit{QeSe} protocol, the security objectives and the adversary model using applied pi-calculus \cite{blanchet2001efficient}. The applied pi-calculus modelling is used as an input to the ProVerif tool. ProVerif then proves if the claimed security properties are fulfilled. We refer the readers to Appendix \ref{app:OS} for more details about the applied pi-calculus semantics and for more details about protocol modelling and the property verification. 
%
%
%
The formal specification includes the following components:

 \textbf{1. Agent model.} \hspace{0.25cm}  Agents represent the protocol parties which execute the roles of the protocol. For instance, we have the sender  and the receiver roles in the protocol. Therefore, in each protocol, each agent performs one or more roles. The agent model is based on a ``closed world assumption'', which means that honest agents show no behaviour other than the one described in the protocol specification (this model corresponds to honest-but-curious threat model in Section \ref{sec:TM}).
\\

\textbf{2. Communication channel.} \hspace{0.25cm}  The communication model describes how the messages are exchanged between the agents. The channel can be private or public according to the threat model. In our scenario the channel is public.
\\

 \textbf{3. Threat model.} \hspace{0.25cm}  Adversaries are modelled as agents that aim to violate the security objectives. ProVerif uses the standard Dolev-Yao adversary model \cite{dolev1983security}. In this model the adversary has complete control over the public communication network.
\\

\textbf{4. Protocol specification.} \hspace{0.25cm} The protocol specification describes the behaviour of each of the roles in the protocol. The system does not execute the actual protocol but it executes the protocol roles performed by the agents. 

To model \textit{QeSe} protocol, we model  Yao's garbled circuit protocol  to verify \textit{Data confidentiality and Fairness}.
Two agents are used to model our protocol in the applied pi-calculus (i.e., one $\p$ and one broker B). 
Based on Yao's circuit two algorithms are defined (namely garble $\garble$ and evaluate $\evall$).  
$\garble$ takes an input function with $n$ bits and outputs a garbled function and a pair of labels for each input bit of the function. $\eval$ takes the garbled function and garbled inputs and returns the required function output.

\par
We model the $\p$'s encryption function as a public parameter represented by a free variables $z_{f}$ (i.e., free variable is similar to global scope in programming languages; that is, free names are globally known).
The private parameters of the  $\p$ and $\broker$ are the protocol inputs (the encryption key and the keyword, respectively). The $\p$'s and $\broker$'s private parameters are represented as free variables $x_\p$ and $y_\broker$, respectively. 

The goal of $\broker$ is to obtain the encryption of its keyword using the $\p$'s key. 
To achieve this we use the protocol specification shown in Figures \ref{fig:token1}. 
The garbling function is modelled using a generated random key, named as the garbling key $\garble(., \key_\garble)$. The garbling key is used to securely garble the circuit at each protocol run (as stated earlier, to ensure the garbled circuit security, $\broker$ receives a fresh, re-encrypted garbled circuit for each input $\garble(y_\broker, \key_\garble)$). 

To model the oblivious transfer, the Broker $\broker$ generates a commitment of each keyword with a fresh generated nonce. The commitment of each keyword cannot be modified and is hidden using the nonce. The commitment is used to request a garbling of the keyword from the $\p$ without disclosing the keyword to the $\p$. We do model the encryption output as  $\evall(z_f, x_\p, y_\broker)$, where both $\p$ and $\broker$ want to find the keyword encryption output without disclosing $x_\p$ to $ \broker$ and $y_\broker$ to $\p$. \\

\textbf{5. Security properties.}  \hspace{0.25cm} These properties specify the security requirements of the protocol defined in Section \ref{sec:Req}.

	\begin{enumerate}
	\item \textit{\textbf{Data Confidentiality:}} Modelling strong secrecy to verify the encrypted SLA's confidentiality is straightforward and modelled easily in ProVerif (adversary learns nothing about the SLA). 
However, to prove the confidentiality property in \textit{QeSe}, we have to prove that the only leakage about the input of the two parties ($\p$ and $\broker$) should come from the result of the evaluated function. 
In other words, if a party obtained the output of the evaluated function, no leakage should occur about the two parties' inputs. 
	
This is proved using the indistinguishability property (the notion of indistinguishability is generally named observational equivalence in the formal model \cite{blanchet2001efficient}). Intuitively, two processes ($P_1$ and $P_2$ are observationally equivalent (i.e., written $P_1\approx P_2$), when an attacker cannot distinguish between the two. Formal definitions of the indistinguishability property can be found in \cite{mobile, automated}. For example, the privacy property of an electronic voting protocol is expressed as \cite{blanchet2001efficient}:
\begin{flalign*}
	P({\sk_A}, v_1) | P({\sk_B}, v_2) \approx P({\sk_A}, v_2) | P({\sk_B}, v_1)
\end{flalign*}

$P$ is the voting process, and an attacker cannot distinguish between the two situations; i) in which $A$ votes for $v_1$ and $B$ votes for $v_2$ from ii) that $A$ votes for $v_2$ and $B$ for $v_1$, where $v_1$ and $v_2$ are the candidates for whom $A$ and $B$ vote. 

To prove the data confidentiality, we model two processes that are replicated an unbounded number of time and executed in parallel. In the first process $P_1$, $\p$ and $\broker$ sends two inputs, represented as free variables $x^0_\p$ and $y^0_\broker$, respectively. While in the second process $P_2$, the $\p$ and $\broker$ sends free variables $x^1_\p$ and $y^1_\broker$, respectively. If the two defined processes are observationally equivalent ($P_1$ $\approx$ $P_2$), then we say that the attacker cannot distinguish between the two process input.  
\\

\textbf{\textit{Theorem 1:}}  Data confidentiality is preserved in the system if $P_1$ and $P_2$ are observationally equivalent ($P_1$ $\approx$ $P_2$). In our protocol this is  proven using ProVerif.\\

\item \textit{\textbf{Fairness:}} In \textit{QeSe}, fairness is preserved if the final result obtained by both parties $\p$ (sender) and $\broker$ (receiver) after a secured computation is consistent. None of the parties can learn the result first and then abort. 
This can be proved using the ProVerif's correspondence property \cite{blanchet2002secrecy}. An example of the correspondence property is, $e(M_1,\ldots,M_j) \Longrightarrow e\textquoteright(N_1,\ldots,N_k)$ , where for any trace of the protocol for each occurrence of event $e(M_1,\ldots,M_j)$, there is a previous occurrence of $e\textquoteright(N_1,\ldots,N_k)$. 
	

	Therefore, we define the fairness property as follows: 
	\begin{flalign}
	\label{eq:int}
	&\p term(z_f,x_\p,y_\broker,r) \Longrightarrow \cr &y_\broker=\com(y_1,n) \wedge r=\evall(z_f,x_\p,y_1)
        \text{ and} \cr
	&\broker term(z_f,x_\p,y_\broker,r\textquoteright) \Longrightarrow \cr &x_\p=\garble(x_1,\key_\garble) \wedge r\textquoteright=\evall(z_f,x_1,y_\broker)
	\end{flalign}

Note that, the $\p$'s and $\broker$'s private parameters are represented using free variables $x_\p$ and $y_\broker$ as stated earlier. The property in equation \ref{eq:int} states that the $\p$ with arguments $z_f, x_\p, y_\broker, r$ terminates a protocol run with $\broker$ after (i) $\p$ receives a commitment of \broker's input with a random value $n$ $(y_\broker=\com(y_1,n))$ and (ii) finding the result $r$ of the of the function $z_f$ on inputs $x_\p$ and $y_1$. Similarly, $\broker$ terminates the protocol run with the $\p$ after receiving the $\p$'s garbled input $x_\p$ and finding the result $r\textquoteright$ of the garbling function $z_f$ on inputs $x_1,y_\broker$.

Using the correspondence property defined~in~equation~\ref{eq:int}, the process $^^21(\p(z_f,x_\p))|^^21(\broker(z_f,y_\broker))$, which models the  two agents processes and  can be executed any number of sessions. 
\\
\\
\textbf{\textit{Theorem 2:}} Fairness of all  possible executions of sessions of honest parties in process $P$ is preserved if the correspondence property defined in equation \ref{eq:int} is true. For our protocol this is proved using Proverif.\\

\end{enumerate}
\end{enumerate}

\subsection{Further Security Considerations}
\label{sec:valid}

Our system and its protocol are designed and verified against honest but curious entities. However, with minor protocol modifications our system mitigate against malicious parties.

\subsubsection{Malicious entities}
\begin{enumerate}[{-}]
\item \textbf{Malicious CSP:}  In the  case of a malicious CSP, it might send (i) an incorrect garbling function to broker $\broker$, and/or (ii) a faulty encrypted secSLA. To prevent  the first form of attack (i), $\broker$ can prove that the garbling is correct by using the ``cut-and-choose'' technique \cite{lindell2007efficient,huang2013efficient}. 
Under this technique, a $\p$ constructs $n$ versions of the circuit, each structured identically but garbled differently so that the keys for each gate in each circuit is unique. Additionally, the $\p$ generates a commitment for each of its garbled inputs. The $\p$ then sends each of the garbled circuits with its garbled inputs to $\broker$. Further, $\broker$ selects $n-1$ versions of the circuit to verify. The $\p$ de-garbles each of the $n-1$ selected circuits, so that $\broker$ can verify that each of the revealed circuits are constructed correctly and as expected. 

$\broker$ checks whether the $\p$'s garbled inputs match their corresponding (previously sent commitments). If everything is correct, $\broker$ evaluates the rest of the circuits and derives the output from them. Thus, if a malicious $\p$ constructs the circuits incorrectly, $\broker$ will detect this with high probability. 
Regarding faulty SLA's, as  explained earlier, an auditor (whose reputation depends on its trustworthiness)  first ensures the validity of the $\p$'s input and thus prevents any CSP from providing faulty encrypted SLAs. 

\vspace{0.5cm}

\item \textbf{Malicious customer:} A single customer   or  a group of colluded customers can try to use the \textit{QRES} system to find each $\p$s security posture. Each of the colluded entities can  specify different requirements and at the end, each customer gets the best matching $\p$'s secSLA according to her/his requirements. To better illustrate  this kind of attack, we show an empirical validation of the proposed system through the secSLA information used in the implementation section; \textit{a secSLA with 50 SLOs, where each SLO is composed of four service levels}. The number of possible SLA combinations is ``$4^{50}$'' (i.e., generally $x^y$, where $x$ is the number of service levels and $y$ is the number of SLOs). Each customer keyword takes on average ``0.52'' seconds to search over one CSP's encrypted secSLA with 50 SLOs with 1 level each (i.e., we demonstrate the system evaluation and performance in Section \ref{sec:imp}; as depicted in Figure \ref{fig:SLOperkeywords}). Thus to find all the possible combinations, it takes $2 * 10^{22}$ years to learn a CSP encrypted secSLA. Hence, this type of attack is not feasible in our system. In addition, countermeasures with specific time between sequential queries could be implemented in the either the broker or the CSP.


\vspace{0.5cm}

\item \textbf{Malicious broker:} A theoretical attack that a broker could perform would be to attempt and alter $\cust$'s keywords.
However, such an attack would be immediately detected by the customer as  at the end they get to know the security details they `negotiated' with the cloud provider.
\end{enumerate}

\subsubsection{Providing CSP Anonymity}

\textbf{System Changes}  \hspace{0.25cm} The \textit{QRES} system should prevent malicious customers to ascertain information about each CSP offered services (i.e., by performing several searching iterations, each iteration with different requirements or by performing frequency analysis). To prevent malicious customers from learning information about CSPs SLAs, \textit{QRES} allows the customers to search for their requirements over ``anonymous'' CSPs encrypted secSLAs. The CSPs anonymity mitigates the malicious customers attacks as customers learn the result of the searching queries and nothing else. 

In order for \textit{QRES} to provide anonymity for the CSPs, the auditor should first generate a unique authentication secret for each CSP during their registration phase. This secret then is send to the broker. Every communication between the broker and the CSP runs over onion routing \cite{reed1998anonymous} anonymity network to ensure anonymity of CSPs. In addition, the CSP generates an authentication challenge using the hash of a nonce and their unique authentication secrets. The challenge together with the nonce are send to the broker. The broker stores the nonces and the challenges alongside with every CSP's secSLA, but cannot tell the real identities of the CSPs. After  finding the best matching CSP, the selected CSP's authentication secret is sent to the auditor by the broker in order to identify the CSP's identity. Next, the auditor sends the selected CSP's identity to $\broker$ to manage the agreement between both the CSP and the customer.
\vspace{0.5cm}

\noindent
\textbf{Verifying  Anonymity}  \hspace{0.25cm}
In order to verify the anonymity property of this modification of the \textit{QRES} system, we use the ProVerif tool.
We model two processes that are replicated an unbounded number of time and executed in parallel. In each process two $\p$s participate by sending their tokens. First, each $\p$ constructs an OR circuit and sends the onion data ($\p_1 \leftrightarrow N_1 \leftrightarrow N_2 \leftrightarrow N_3$) and ($\p_2 \leftrightarrow N_1 \leftrightarrow N_2 \leftrightarrow N_3$). Then, each of the intermediate nodes ($N_1$ and $N_2$) removes one layer of encryption and at the end forwards the onion to $N_3$. Finally, once the exit node $N_3$ receives the two onions from the two $\p$s, it removes the last layer and sends the messages to $\broker$.

In the first process $P_1$, $\p_1$ and $\p_2$ sends two tokens $t_1$ and $t_2$, respectively. Once the exit node $N_3$ removes the last onion layer, it sends the message to $\broker$ on a public channel ($t_1 || t_2$). In the second process $P_2$, the two tokens are swapped; such that $\p_1$ and $\p_2$ sends $t_2$ and $t_1$, respectively. Similarly, $N_3$ publishes the message on a public channel ($t_2 || t_1$). The $\p$s anonymity is preserved if an attacker cannot distinguish between the two messages and hence, cannot learn which token is sent by which $\p$. Nodes transfer messages to each other using a  public channel.

If the two defined processes are observationally equivalent ($P_1$ $\approx$ $P_2$), then we say that the attacker cannot distinguish between $t_1$ and $t_2$, which means the attacker is unable to distinguish when the message changes. Hence, the attacker is not able to link two communication streams of the same $\p$, and thus cannot learn which message is sent by which $\p$.  

	\textbf{Theorem 3:} The observational equivalence of $P_1$ and $P_2$ defined in equation \ref{eq:in} holds ($P_1$ $\approx$ $P_2$).	
		
\begin{flalign}
		\label{eq:in}
		P_1(t_1, t_2) \approx P_2(t_2, t_1)
		\end{flalign}

%% file: src/10-Implementation.tex

\section{Implementation and Evaluation}
\label{sec:imp}

\noindent
\textbf{Implementation.}\hspace{0.25cm}
\textit{QRES} is implemented in Java, using Apache-Tomcat 9.0 and Amazon DynamoDB \cite{ddb} database, which is a NoSQL database service. 
The DynamoDB supports both string and key value store models which we used to store the encrypted tokens ($\broker$ side). 
We used Tomcat web server on our Ubuntu machine to transmit  the encrypted data to Amazon DynamoDB (the $\p$ side). 
We implement the hash functions using HMAC based on SHA-256. 
The symmetric encryption scheme is implemented based on AES 128. 
Moreover, we use $\beta$ = 128 for nonces and $8$ bytes per token. 
Furthermore, we modified Yao's garbled circuit coding \cite{git} to fit our encryption scheme. 
Finally, we implemented two Java Server Page (JSP) files to send the tokens and to search for the keywords. 
The complete source code along with a detailed explanation of setting up and using the \textit{QRES} can be found at \cite{qres}. 

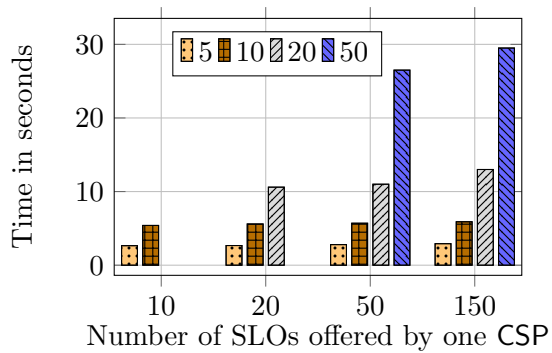
\begin{figure}[htb]
\centering
\begin{tikzpicture}
\begin{axis}[
ybar,
	ybar=2pt,
	bar width=6pt,
width=7cm, height=5cm, enlargelimits=0.15,
legend style={at={(0.4,0.95)},
anchor=north,legend columns=-1.5},
ylabel={Time in seconds},
xlabel={Number of SLOs offered by one $\p$},
symbolic x coords={10,20,50,150},
grid=major,
xtick=data,
nodes near coords align={vertical},
error bars/.cd,
 legend image code/.code={%
      \draw[#1] (0cm,-0.1cm) rectangle (0.15cm,0.2cm);}
]
\addplot 
[black,fill={rgb:orange,1;yellow,2;pink,5},postaction={pattern=dots}]
coordinates {(10,2.65) (20,2.67) (50,2.795)(150,2.92)};
\addplot 
[black,fill={rgb:red,4;green,2;yellow,1},postaction={pattern=grid}]
coordinates {(10,5.4) (20,5.6) (50,5.7)(150,5.9)};
\addplot 
[black,fill={rgb:black,1;white,6},postaction={pattern=north east lines}]
coordinates {(20,10.6) (50,11)(150,13)};
\addplot 
[black,fill=white!40!blue,postaction={pattern=north west lines}]
coordinates {(50,26.5)(150,29.5)};
\legend{$5$, $10$, $20$, $50$}
\end{axis}
\end{tikzpicture}
\caption{Time used by a customer to search for her/his different keywords over varied number of SLOs offered by one CSP.}
  \label{fig:SLOperkeywords}
\end{figure}

\vspace{0.5cm}
\noindent
 \textbf{Performance Evaluation.}\hspace{0.25cm} We evaluate the performance of the  \textit{QeSe} protocol running between the $\p$ and the $\broker$ using two use-cases. First, we use \textit{QReS} with  one $\p$ which is offering various tokens (SLOs) in its secSLA . Note that, the $\p$ is offering \textit{one service level for each SLO} in its secSLA. Each token (8 bytes) is encrypted and sent to Amazon DynamoDB.    
 Figure \ref{fig:SLOperkeywords} shows the amount of time spent by a broker to search for the customer keywords (i.e. 5,10, 20 and 50 keywords ) over 10, 20, 50 and 150 SLOs provided by one $\p$. The use-case shows that the time  to search for the customer's different keywords over various number of SLOs is almost the same despite the number of the offered SLOs. This is expected as the most time consuming part of the computation, that is the computation of the circuits, is the same despite the amount of offered SLOs.

\begin{figure}
\centering
\begin{tikzpicture}
\begin{axis}[
width=7.5cm, height=5.5 cm,
xmin=0, xmax=32,
ymin=0, ymax=2700,
xtick={1,10,30},
ytick={100,500,1000,1500, 2000},
xlabel={Number of CSPs' secSLAs each with150 SLOs},
ylabel={Time in seconds},
grid=major,
legend pos=north west
]
\addplot[
    color=blue,
    mark=*,
    line width=1.5pt,
    mark size=2.5pt,
    ] table {Q1.dat};
\addplot[
    color=red,
    mark=square*,
    line width=1.5pt,
    mark size=2.5pt,
    ]  table {Q2.dat};
\addplot[
    color=gray,
    mark=*,dashed,
    line width=1.5pt,
    mark size=2.5pt,
    ]  table {Q3.dat};
\legend{20 keywords,50 keywords,150 keywords}
\end{axis}
\end{tikzpicture}
\caption{Time used by the customers to search for their different keywords over 150 SLOs offered by varied CSPs.}
  \label{fig:perfComp}
\end{figure}
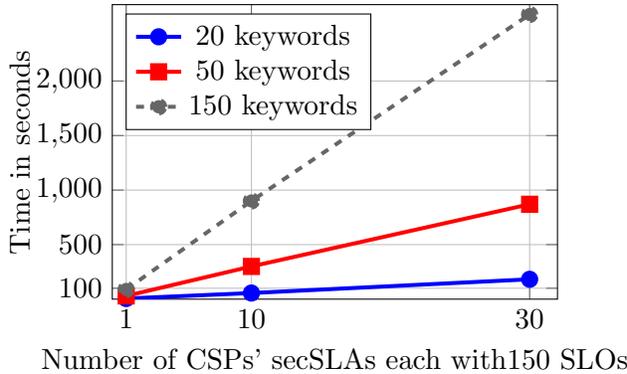

Further, we explore the case in which a customer compares  various CSPs based on their advertised secSLAs.
Figure \ref{fig:perfComp} shows the amount of time a broker spends to search for the customer requirements (keywords) over 1, 10 and 30 CSPs' secSLAs where each secSLA consists of 150 SLOs. For each CSP's secSLA, we extracted 150 tokens (8 bytes each) which are then encrypted and sent to Amazon DynamoDB (in our current implementation we saved each CSP token in a different table). The use-case shows a linear time  progression as the broker searches for more  keywords in each $\p$'s table.

The SLOs are extracted from the public STAR repository \cite{csa_star}. The rationale for using STAR repository is that (i) to the best of our knowledge no other cloud SLA repositories are publicly available and (ii) major CSPs are still in the process of restructuring their SLAs by leveraging the recently published ISO/IEC 19086. Currently, the STAR contains reports with $\p$'s answers to Consensus Assessments Initiative Questionnaire \cite{csa_caiq} with yes/no answers. Furthermore, we utilized other requirements defined in multiple research projects such as A4cloud \cite{A4Cloud}, CUMULUS \cite{Cumulus}, and SPECS \cite{Specs}. 


%% file: src/11-Relatedwork.tex

\section{Related Work}
\label{sec:sota}
In this section we describe the related work which falls into four categories: security quantification, secured computation, searchable encryption, and formal analysis.
\vspace{0.5cm}
\noindent
\textbf{Security quantification:}\hspace{0.25cm}  Security requirements for non-cloud scenarios have been addressed by Casola et al. \cite{casola2006sla}, who propose a methodology to evaluate security SLAs for web services. In~\cite{casola2006sla}, the authors propose a methodology to assess security SLAs for web services. In~\cite{frankova2007service} and \cite{krautsevich2011general}, the authors propose a technique to aggregate security metrics from web services security SLAs. However, in contrast to our work, the previous work did not empirically validate the specified metrics. 
In \cite{almorsycollaboration}, the authors propose the notion of evaluating cloud SLAs by introducing a metric to benchmark the security of a CSP based on categories. However, all of the previously mentioned approaches do not allow the CSPs to specify their security posture. 

\vspace{0.5cm}
\noindent
\textbf{Secure computation:}\hspace{0.25cm}  Fully homomorphic encryption (FHE) \cite{gentry2009fully} and other general functional encryption \cite{garg2016candidate} schemes can be used to compute functions over encrypted data. However, they do not address all our desired security properties, and they are prohibitively slow for the selected usecase. Some recent systems such as CryptDB \cite{po2011} and Mylar \cite{popa2014} support secure computation efficiently. However, these systems enables certain types of search over the encrypted data which  are not matching our required properties. In \cite{sherry2015blindbox}, the authors utilize secure computation on encrypted data for deep packet inspection.  However, the existing two part computation over encrypted data schemes do not provide the desired property and functionality of \textit{QRES}. 

\vspace{0.5cm}
\noindent
\textbf{Searchable encryption:}\hspace{0.25cm}  There exist different kinds of searchable encryption schemes named deterministic \cite{bellare2007deterministic} and randomized \cite{song2000practical,kamara2012dynamic}. In \cite{sherry2015blindbox}, Sherry et al. introduced an encryption scheme that achieves both the detection speed of the deterministic encryption and the security of the randomized encryption.

\vspace{0.5cm}
\noindent
\textbf{Formal analysis:}\hspace{0.25cm}  In \cite {backes2013introducing}, the authors provide practical repudiation for autonomous communication networks by tracing back the selected outbound traffic to the predecessor node. They conduct a formal security analysis of the OR protocol using ProVerif by formalizing anonymity and no forward traceability as observational equivalence relations, and backward traceability and no false accusation as trace properties. In \cite {backes2008zero} an abstraction of non-interactive zero-knowledge proofs within the applied pi-calculus is presented. The authors transform their abstraction into an equivalent formalization that is accessible to ProVerif. The authors in \cite{blanchet2008automated} study the formal security properties of well-established protocols for secure file sharing on untrusted storage. The protocol modeling and properties verification are studied using the automatic protocol verifier ProVerif.

%% file: src/12-conclusion.tex

\section{Conclusion}
\label{sec:conc}

Cloud Service Providers are a lucrative target because of the amount of data they process. A  disruption of their services due to a hack, can cause them a terrible financial loss and many problems for the customers whose data are compromised. The inclusion of security implementations information in the Service Level Agreements is risky, as it would enable malicious entities to better orchestrate their attacks and easier discover  vulnerabilities. Hence, many providers do not include them in their SLAs. The SLAs however, are the agreement between customers and  providers for the services offered, and the security implementations is an important factor for choosing a cloud provider.

We tackle this problem with a system called \textit{QRES}. Our system enables CSPs to create security SLAs and publish them encrypted. With the help of an intermediate node, customers can find the cloud provider better matching their security needs by contrasting their requirements against the encrypted secSLAs.
Our system is utilizing  the \textit{QeSe} protocol, a searchable encryption scheme protected by secure two party computation. The inputs of every party  remain private while the output is only learned by the broker and consequently the customer.  

We implement \textit{QRES} and formally verify its security and privacy properties using ProVerif. In our real word tests, using existing standardized SLAs by the latest industrial standard, the system proved to be fast for the required use case. In our measurements, it requires less than 30 seconds to privately search for 50 customer keywords at a CSP.

\textit{QRES} is the first step towards  providing  security assurance and ``transparency" between cloud customers and CSPs, while  at the same time ensures the confidentiality of the CSPs sensitive information. Moreover, we aim to extend the system model presented in this paper  by checking whether general functional encryption \cite{garg2016candidate} schemes can address all the desired security properties of our model.

%% file: src/notation-table.tex
\begin{table*}[t]
\centering
     \caption{Notation and Preliminaries}
     \label{table:terms}    
  \begin{tabular}{| p{3cm} | p{10.5cm}| }
  \hline
  \textbf{Term} & \textbf{Definition}\\\hline
\hline  
\slo/ &Service Level Objective (SLO)\\\hline
  \cust & Customer\\\hline
  \broker & Broker\\\hline
  	$m$ & Plaintext\\\hline
  	$c$ & Ciphertext\\\hline
     \key & Symmetric encryption/decryption key.\\\hline
  	 $\pk_{\x}$ & Public key of party $\x$.\\\hline
     $\sk_{\x}$  & Private key of party $\x$.\\\hline
     $\cert_{\x}$ &X.509 certificate of party $\x$.\\\hline
     $\kxvauth$ & Authentication secret generated and sent to party $\x$.\\ \hline
     $\nonce_{\x} \in \{0,1\}^{\beta}$ & Random number generated by party $\x$ from the set of all binary strings of length ${\beta}$. \\ \hline
    $y \sample \alg(x)$ & On input $x$, algorithm $\alg$ binds the output to variable $y$.\\\hline
	$m_1||m_2$ & Concatenation of $m_1$ and $m_2$.\\  \hline   
     $\kgen$ & key generation.\\\hline
     $\enc$ & Encryption algorithm\\\hline 
     $\dec$ & Decryption algorithm \\\hline 
     $c \sample \enc(\pk,m)$ & Takes as input a public key $\pk$, a plaintext $m$, and outputs a ciphertext $c$.\\\hline
  $m\sample \dec(\sk,c)$ & Takes as input a secret key $\sk$, a ciphertext $c$, and outputs a plaintext $m$.\\\hline
  $s_m \sample \sig(\sk,m)$ & Signs $m$ with a secret key $\sk$, and outputs a signed message $s_m$.\\\hline
   $b \sample\vrfy(\pk,s_m)$ & Verifies the signature using the public key $\pk$ and returns $1$ if the signature is valid for the given message and $0$ if not. 
\\\hline
$y\sample\hash(x)$ & Hash function which takes as input $x$, and responds with a random outputs $y$. Ideally, a hash function is defined as a random oracle that responds with a random output $y \in \y$ to each given input for $x \in \x$. Where $\x$ is the set of possible messages, $\y$ is a finite set of possible digests.  \\\hline
%
%
%
$t^1,\ldots, t^n$ & The SLA XML file can be split into $n$ substrings. For example, the substrings generated from Listing 1.1 are: ``$level3 || 3$'' and ``$level2 || 4$'' (cf., Section \ref{sec:Token})\\\hline 
$t^1_{\p_1},\ldots, t^n_{\p_1}$ & Substrings generated by $\p_1$ and are named tokens, where $n$ is the number of tokens\\\hline
$w^1_{\cust},\ldots, w^n_{\cust}$ & Substrings generated by the customer $\cust$ and are named keywords, where $n$ is the number of keywords\\\hline
  \end{tabular}
\end{table*}

%% file: src/13-app.tex

\section*{Appendix}

\section*{Security Definition}
\label{app:secdef}
%
 A function $\nu()$ is negligible in an input parameter $k$ if for every positive polynomial $p()$ and all large $k$, $\nu(k) \leqslant 1/p(k)$. An Encryption Scheme $ES$ is a tuple of polynomial time algorithms $(\kgen, \enc, \dec)$ where: 
\begin{enumerate}[-]
\item $\key \sample \kgen (\secparam)$: It takes as input a security parameter \secparam and outputs a secret key $\key$.
\item $c \sample \enc(\pk,m)$: It takes as input a public key $\pk$, a plaintext $m$, and outputs a a ciphertext $c$.
\item $m\sample \dec(\sk,c)$: It takes as input a secret key $\sk$, a ciphertext $c$, and outputs a a plaintext $m$.
\end{enumerate}
Note that, for symmetric encryption scheme $\pk=\sk$. We denote $s \sample \sig(\sk,m)$ for signing $m$ with a secret key; verify the signature using the public key $\verify(\pk,s)$ and recover $m$. 

A Symmetric searchable encryption $SSE$ allows an entity to encrypt data in such a way that it can later generate search tokens to search over the encrypted data and find the required files only. 
Based on the $SE$ schemes definition in \cite{kamara2012dynamic,sherry2015blindbox}, we define an $SSE$ scheme as a tuple of polynomial time algorithms $(\kgen$, $\enc$, $\keyenc$, $\search)$ where:
\begin{enumerate}[-] 
\item $\key \sample \kgen (\secparam)$: It takes as input a security parameter \secparam and outputs a secret key $\key$.
\item $c^{t^1},\ldots,c^{t^n} \sample \enc(\key,t^1,\ldots, t^n)$: It takes as input a secret key $\key$ and a set of $n$ substrings, and outputs a set of $n$ encrypted substrings.
\item $c^w \sample \keyenc(\key,w)$: It takes as input a secret key $\key$ and one substring, and outputs an encrypted substring.
\item $d \sample \search(c^w,c^{t^1},\ldots,c^{t^n})$: It searches for the encrypted substring from $c^{t^1},\ldots,c^{t^n}$ which matches $c^w$. It takes as input $n+1$ encrypted substrings and outputs the matched encrypted substring index. If there is no any encrypted substring matching $c^w$, it outputs $\perp$. 
\end{enumerate}

\section*{Operational semantics}
\label{app:OS}
The rules that define the operational semantics of applied pi-calculus and ProVerif are adapted from \cite{blanchet2009automatic}. The identifiers $\vara, \varb, \varc, \vark$ and similar ones range over names, and $\varx, \vary$ and $\varz$ range over variables. As detailed in \cite{blanchet2009automatic}, set of symbols is also assumed for constructors and destructors such that $f$ for a constructor and $g$ for a destructor. Constructors are used to build terms. Therefore, the terms are variables, names,
and constructor applications of the form $f(M_1, \ldots , M_n)$.

We use the constructors and destructors defined in \cite{blanchet2009automatic} as an initial step to represent the cryptographic operations as depicted in Figure \ref{fig:con}. We added other different constructors/destructors which are used to define our protocol. Constructors and destructors can be public or private. The public ones can be used by the adversary, which is the case when not stated otherwise. The private ones can be used only by honest participants.

\begin{figure}[!ht]
\begin{flushleft}
\textbf{Symmetric enc/dec:}\\
Constructor: encryption of $\varx$ with the shared secret key $\vark$, $\senc (\varx, \vark)$\\
Destructor: decryption $\sdec(\senc(\varx,\vark),\vark) \rightarrow \varx$

\textbf{Asymmetric enc/dec:}\\
Constructor: encryption of $\varx$ with the public key generation from a secret key $\vark$, $\pkk$, $\aenc (\varx, \pkk)$\\
Destructor: decryption $\adec(\aenc(\varx,\pkk),\vark) \rightarrow \varx$

\textbf{Signatures:}\\
Constructors: signature of $\varx$ with the secret key $\vark$, $\sgnm( \varx, \vark)$\\
Destructors: signature verification using he public key generation from a secret key $ \vark, \pkk, \vgnm(\sgnm( \varx, \vark), \pkk) \rightarrow \varx$

\textbf{One-way garbling function:}\\
Constructors: garbling of $\varx$ with the key $\vark, \garbled(\varx,\vark)$

\textbf{Evaluation function:}\\
Constructors: evaluation function of garbling of variables $\varx$, $\vary$, and $\varz$ with the key $\vark$, $\evaluated(\garbled(\varx,\vark),\garbled(\vary,\vark),\garbled(\varz,\vark))$

\textbf{Commitment:}\\
Constructors: committing $\varx$ with a fresh nonce $\varn$ $\vark, \commitd(\varx,\varn)$

\caption{Constructors and destructors}
\label{fig:con}
\end{flushleft}
\end{figure}

The operational semantics used are presented in Figure \ref{fig:sem}. A semantic configuration is a pair {\names,\procs} where the {\names} is a finite set of names and {\procs} is a finite multiset of closed processes. The semantics of the calculus is defined by a reduction relation $\rightarrow$ on semantic configurations as shown in Figure \ref{fig:sem}. 
The process $\ev(M).\proc$ executes the event $\ev(M)$ and then executes $\proc$. The input process $in(M,\varx).\proc$ inputs a message, with $\varx$ bound to it, on channel $M$, and executes $\proc$. The output process $out(M,N).\proc$ outputs the message $N$ on the channel $M$ and then executes \proc. 
\begin{figure*}
\fbox{
\pseudocode{
(Nill) \hspace{0.3cm} \names,\procs \cup \{0\} \rightarrow \names,\procs \\
(Repl) \hspace{0.3cm} \names,\procs \cup \{!\proc\} \rightarrow \names,\procs \cup \{\proc,!\proc\} \\
(Par) \hspace{0.3cm} \names,\procs \cup \{\proc | \q\} \rightarrow \names,\procs \cup \{\proc,\q\} \\
(Par) \hspace{0.3cm} \names,\procs \cup \{\proc | \q\} \rightarrow \names,\procs \cup \{\proc,\q\} \\
(New)  \hspace{0.3cm} \names,\procs \cup \{ ({\pcnewd a}) \proc \} \rightarrow \names \cup \{ {a}' \},\procs \cup \{ \proc \{ {a}' / {a}\} \}  \text{  where  } a' \notin \names\\
(I/O)  \hspace{0.3cm}  \names,\procs \cup \{out(c,M).\q, in(c,\varx).\proc\} \rightarrow  \names,\procs \cup \{ \q, \proc \{M / \varx \}\}\\
(Cond1) \hspace{0.3cm} \names,\procs \cup \{ \text{if } M = N \text{ then } \proc \text{ else } \q \} \rightarrow \names,\procs \cup \{ \proc \} \text{ if } M = N\\
(Cond2) \hspace{0.3cm} \names,\procs \cup \{ \text{if } M = N \text{ then } \proc \text{ else } \q \} \rightarrow \names,\procs \cup \{ \q \} \text{ if } M \neq N\\
(Let) \hspace{0.3cm} \names,\procs \cup \{ \text{let } \varx = g(M_1, \ldots, M_n) \text{ in } \proc \text{ else } \q \} \rightarrow \names,\procs \cup \{ \proc \{M'/\varx \}\} \\ \text{ if } g(M_1, \ldots, M_n) \rightarrow M'\\
}
}
\caption{Operational semantics \cite{automated}}
\label{fig:sem}
\end{figure*}
The nil process $0$ does nothing. The process $\proc | \q$ is the parallel
composition of $\proc$ and $\q$. The replication $!\proc$ represents an unbounded number of copies of $\proc$ in parallel. $({\pcnewd a}) \proc$ creates a new name $a$ and then executes $\proc$. The conditional if $M = N$ then $\proc$ else $\q$ executes $\proc$ if $M$ and $N$ reduce to the same term at runtime; otherwise, it executes $\q$. Finally, let $\varx = M$ in $\proc$ as syntactic for $\proc \{M / \varx \}$ which is the process obtained from $\proc$ by replacing every occurrence of $\varx$ with $M$. As usual, we may omit an else clause when it consists of $0$.

\section*{Protocol modelling and properties verification}
\label{app:formal}
In this section we model the \textit{Qese} protocol depicted in Figures \ref{fig:token1}, then verify the fairness property.

\pseudocode{
\small
- \proc_{\broker}(sk_{\broker},pk_{\broker},m_{\broker})=!in(c,m).({\pcnewd b}) \\ event(e_1({\commitd(m_{\broker},b)})).\\out(c,{\commitd(m_{\broker},b)}).
in(c,m''). let ((\varx_{\broker},\varx_f, \varx_{\p},{m_{x}})=\\
\adec(m'',sk_{\broker}))in 
if m_x={\commitd(m_{\broker},b)} then\\
event(e_{\broker}({\commitd(m_{\broker},b)},\varx_{\broker},\varx_f, \varx_{\p},\evaluated(\varx_{\broker},\varx_f, \varx_{\p}))).\\out(c,\evaluated(\varx_{\broker},\varx_f, \varx_{\p}))
\\
- \proc_{\p}(pk_{\broker},m_f,m_{\p})=in(c,m').let((\vary_{\broker}= \garbled(m',\vark)) | \\(\vary_{f}= \garbled(m_f,\vark)) | (\vary_{\p}= \garbled(m_{\p},\vark))) in\\ 
event(e_2(m',\vary_{\broker}, \vary_f, \vary_{\p})).\\out(c,(\senc((\vary_{\broker},\vary_{f},\vary_{\p},m'), sk_{\p}))\\
.in(c,m''')\\ if m'''=\evaluated(\vary_{\broker},\vary_f, \vary_{\p}) then\\
event(e_{\p}(m',\vary_{\broker},\vary_f, \vary_{\p}, m'''))
\\
- \proc(\pcnewd m_f)(\pcnewd m_{\p})(\pcnewd m_{\broker})(\pcnewd sk_{\broker}) let pk_{\broker}=pk_{sk_{\broker}}in\\
out(c,pk_{\broker}). \proc_{\broker}(sk_{\broker},pk_{\broker},m_{\broker}) | \proc_{\p}(pk_{\broker},m_f,m_{\p})\\
}

The channel $c$ is public so that the adversary can send, replay and get any messages sent over it. We use a single public channel and not two or more channels because the adversary could take a message from
one channel and relay it on another channel, thus removing any difference between the
channels. The process $\proc$ begins with the creation of the secret and public keys of $\broker$, and the creation of messages $m_f , m_{\p} , m_{\broker}$ The public key is output on channel $c$ to model that the adversary has it in its initial knowledge. Then the protocol itself starts: $\proc_{\broker}$ represents $\broker$, $\proc_{\p}$ represents the $\p$. Both principals can run an unbounded number of sessions, so $\proc_{\broker}$ and $\proc_{\p}$ start with replications.

We consider that $\broker$ first inputs a message containing the encrypted tokens and then starts the protocol run by choosing a nonce $b$, and executing the event $e_1({\commitd(m_{\broker},b)})$, where $m_{\broker}$ is initially added to the $\broker$ knowledge. 
Intuitively, this event records that $\broker$ sent $Message_1$ of the protocol. Event $e_1$ is placed before the actual output of $Message_1$; this is necessary for the desired correspondences to hold: if event $e_1$ followed the output of $Message_1$, we would not be able to prove that event $e_1$ must have been executed, even though
$Message_1$ must have been sent, because $Message_1$ could be sent without executing
event $e_1$, as stated in \cite{blanchet2009automatic}. The situation is similar for events $e_2 , e_{\broker}$ and $e_{\p}$. 

Next, $\broker$ receives the garbling of $\p$'s inputs as well as the garbling of the committed messages encrypted with its public key. $\broker$ decrypts the message using its secret key $sk_{\broker}$. If decryption succeeds $\broker$ checks if the message has the right form using the pattern-matching construct $let ((\varx_{\broker},$ $\varx_f, \varx_{\p},={m_{\broker}})=\adec(m'',sk_{\broker}))in$. Then $\broker$ executes the event $e_{\broker}({\commitd(m_{\broker},b)},\varx_{\broker},\varx_f, \varx_{\p},$\\ $\evaluated(\varx_{\broker},\varx_f, \varx_{\p}))$, to record that it has received $Message_2$ and sent $Message_3$
of the protocol. Finally, $\broker$ sends the last message of the protocol $\evaluated(\varx_{\broker},\varx_f, \varx_{\p})$. 

After sending this message, $\broker$ executes some actions needed only for specifying properties of the protocol. When the received message $m_x = \commitd(m_{\broker},b)$, that is, when the session is between $\broker$ and $\p$, $\broker$ executes the event $e_{\broker}({\commitd(m_{\broker},b)},\varx_{\broker},\varx_f, \varx_{\p},$\\$\evaluated(\varx_{\broker},\varx_f, \varx_{\p}))$, to record that $\broker$ ended a session of the protocol, with the participant ($\p$), which is authenticated using the authentication key. $\broker$ also outputs the evaluation function output $\evaluated(\varx_{\broker},\varx_f, \varx_{\p})$. 


The process $\proc_{\p}$ proceeds similarly: it executes the protocol, with the additional event $e_2(m',\vary_{\broker}, \vary_f, \vary_{\p})$ to record that $Message_1$ has been received and $Message_2$ has been sent by $\p$, in a session with the participant of public key $pk_{\broker}$ and the received message $m'$. After finishing the protocol itself, when $m'''=\evaluated(\vary_{\broker},\vary_f, \vary_{\p}) $, that is, when the session is between $\broker$ and $\p$, $\proc_{\p}$ executes the event $e_{\p}(m',\vary_{\broker},\vary_f, \vary_{\p}, m''')$, to record that $\p$ finished the protocol, and outputs $m'''$.

The events will be used in order to formalize fairness. For example, we formalize that, if $\p$ ends a session of the protocol $e_{\p}(m',\vary_{\broker},\vary_f, \vary_{\p}, m''')$, then (a) $\broker$ has started a session of the protocol by committing $m_{\broker}$ with the nonce $n_{\broker}$, and (b) $\p$ outputs the evaluation function $\evaluated(\vary_{\broker},\vary_f, \vary_{\p})$. Furthermore, $\broker$ ends a session of the protocol, then (a) $\p$ has already garbled the $\broker$'s committed input message, and (b) $\broker$ outputs the evaluation function $\evaluated(\varx_{\broker},\varx_f, \varx_{\p})$.

Next, we formally define the correspondences in order to verify the fairness property. We prove correspondences in the form of \textit{if an event $e$ has been executed, then events $e_1,\ldots, e_m$
have been executed}. These events may include arguments, which allows one to relate the values of variables at the various events. We can prove that each execution of $e$ corresponds
to a distinct execution of some events, and that the events have been executed in a certain order. We assume that the protocol is executed in the presence of an adversary that can listen to all messages, compute, and send all messages it has, following the so-called Dolev-Yao model \cite{dolev1983security}. Thus, an adversary can be represented by any process that has a set of public names $Init$ in its initial knowledge and that does not contain events.

As presented in system model, the correspondence event $e_{\p}(\varx_1,$\\$\varx_2, \varx_3, \varx_4,\varx_5)$ $\leadsto e_1(\varx_1) \wedge e_2(\varx_1,\varx_2, \varx_3, \varx_4) \wedge e_{\broker}(\varx_1,\varx_2, \varx_3, \varx_4,\varx_5)$ means that, if the event $e_{\p}(\varx_1,\varx_2,\varx_3, \varx_4,\varx_5)$ has been executed, then the events $e_1(\varx_1), e_2(\varx_1,\varx_2, \varx_3, \varx_4)$ and $e_{\broker}(\varx_1,\varx_2, \varx_3, \varx_4,\varx_5)$ have been executed, with the same value of the arguments $\varx_1,\varx_2, \varx_3, \varx_4,\varx_5$.